%% file: main.tex
\documentclass[sigconf, natbib=true, anonymous=false]{acmart}
\AtBeginDocument{%
  }

\setcopyright{acmlicensed}
\copyrightyear{2026}
\acmYear{2026}
\acmDOI{XXXXXXX.XXXXXXX}
\acmConference[CIKM'26]{the 35th ACM International Conference on Information and Knowledge Management}{November 07--11,
  2026}{Rome, Italy}
\acmISBN{978-1-4503-XXXX-X/2026/06}

\usepackage{kotex}
\usepackage[inkscapelatex=false]{svg}
\usepackage{booktabs}
\usepackage{float}
\usepackage{multirow}
\usepackage{booktabs} 
\usepackage{array}    
\usepackage[table]{xcolor}
\usepackage{tabularx}
\usepackage{booktabs} 
\usepackage[most]{tcolorbox}
\usepackage{enumitem}
\usepackage{balance}

\usepackage{algorithm}
\usepackage{algorithmicx}
\usepackage{algpseudocode}

\algnewcommand\algorithmicinnerfunc{\textbf{Inner function:}}
\algnewcommand\InnerFunc{\item[\algorithmicinnerfunc]}

\usepackage{amsmath, amsthm}
\theoremstyle{definition}
\newtheorem{definition}{Definition}

\theoremstyle{example}
\newtheorem{example}{Example}

\theoremstyle{proposition}

\theoremstyle{corollary}
\newtheorem{corollary}{Corollary}

\begin{document}

\title[GraFine: Retrieval-Time Refinement for Efficient Graph RAG over Corpus Graphs]{GraFine: Retrieval-Time Refinement\\for Efficient Graph RAG over Corpus Graphs}

\author{Seonho An}
\orcid{0009-0008-9280-0254}
\authornote{Co-first author.}
\affiliation{%
  \institution{KAIST}
  \country{Daejeon, Republic of Korea}
}
\email{asho1@kaist.ac.kr}

\author{Chaejeong Hyun}
\orcid{0009-0007-8349-4868}
\authornotemark[1]
\affiliation{%
  \institution{KAIST}
  \country{Daejeon, Republic of Korea}
}
\email{hchaejeong@kaist.ac.kr}

\author{Min-Soo Kim}
\authornote{Corresponding author.}
\orcid{0000-0002-5065-0226}
\affiliation{%
  \institution{KAIST}
  \country{Daejeon, Republic of Korea}
}
\email{minsoo.k@kaist.ac.kr}

\renewcommand{\shortauthors}{An et al.}

\newcommand{\vs}{\ensuremath{\mathcal{O}_{\mathrm{v}}}}
\newcommand{\gs}{\ensuremath{\mathcal{O}_{\mathrm{g}}}}
\newcommand{\model}{\ensuremath{\mathcal{O}_{\mathrm{m}}}}
\newcommand{\vgs}{\ensuremath{\mathcal{O}_{\mathrm{vg}}}}
\newcommand{\gmodel}{\ensuremath{\mathcal{O}_{\mathrm{gm}}}}
\newcommand{\vgg}{\ensuremath{\mathcal{O}_{\mathrm{vgg}}}}
\newcommand{\prn}{\ensuremath{\mathcal{O}_{\mathrm{cprn}}}}
\newcommand{\cmark}{\checkmark}
\newcommand{\xmark}{\ensuremath{\times}} 

\definecolor{pastelgreen}{HTML}{D5E8D4} 
\definecolor{pastelpurple}{HTML}{E1D5E7} 
\definecolor{pastelorange}{HTML}{FFE6CC} 
\definecolor{pastelblue}{HTML}{D6EAF8} 

\definecolor{navyblue}{HTML}{000080}


\begin{abstract}
Graph RAG on corpus graphs enhances retrieval by leveraging intermediate node content as contextual clues to uncover unretrieved oracle nodes.
However, existing methods suffer from two critical blind spots, namely semantically blind graph expansion and topology blind pruning, or else rely on prohibitively slow retrieval and generation interleaving.
To address this, we formalize these limitations through an operational taxonomy and propose a retriever design that couples semantics-\textit{aware} adding with graph-\textit{aware} pruning under efficiency constraints. We instantiate this design as \textbf{GraFine}, which alternates between two refinement stages: Semantic Proximity eXpansion (SPX) for semantics-aware node addition, and a Graph Smoothing Reranker (GSR) for graph-aware pruning.
Experiments on reference networks and text-rich knowledge graphs show that GraFine improves retrieval accuracy and generation quality while maintaining time-efficiency. Furthermore, we introduce Topological Recall~(TR), a metric that quantifies the topological proximity between retrieved and oracle nodes. Our analysis using TR confirms that GraFine more effectively steers refinement toward undiscovered oracle nodes by leveraging intermediate contextual clues.
Our code is available at this link: \href{https://github.com/asmath472/GraFine}{\textcolor{navyblue}{\textbf{github.com/asmath472/GraFine}}}.
\end{abstract}




\maketitle

\input{sec1_intro}
\input{sec2_preliminaries}

\input{sec3_taxonomy}

\input{sec3_methodologies}

\input{sec4_experiments}
\input{sec5_result-and-analysis}

\input{sec6_related-works}
\input{sec7_conclusion}
\input{sec8_genAI}

\bibliographystyle{ACM-Reference-Format}
\bibliography{sample-base}


\end{document}

%% file: sec1_intro.tex
\section{Introduction}
\label{sec:introduction}

Retrieval-Augmented Generation (RAG) has emerged as a widespread solution to mitigate the inherent limitations of Large Language Models (LLMs), such as hallucinations and outdated parametric knowledge~\cite{gao2023retrieval}. 
However, RAG methods that rely primarily on vector search (which we refer to as Vector RAG) often struggle to capture structural dependencies and non-textual relations, such as reference links between documents, because retrieval is largely driven by vector similarity alone~\cite{edge2024local, guo2025lightrag, gutierrez2025from, zhu-etal-2025-knowledge}. 
To address this, recent studies have proposed \textbf{Graph RAG} methods~\cite{guo2025lightrag, jimenez2024hipporag, sun2024thinkongraph, ma2025thinkongraph, gutierrez2025from, chen2025pathrag, huang2025ket, zhu-etal-2025-knowledge}, which incorporate a \textit{graph} structure into the retrieval process to capture relationships via edges~\cite{han2024retrieval}.

\begin{figure*}[t!]
    \centering
    \includegraphics[width=\textwidth]{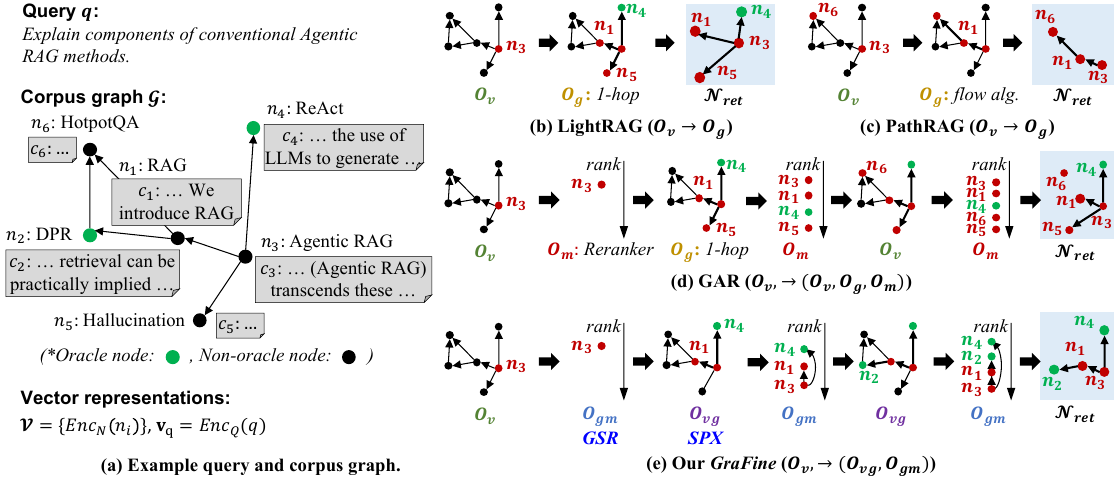}
    \vspace{-0.8cm}
    \Description{Conceptual comparison of graph retrieval workflows based on retrieval operations.
    (a) illustrates the inputs for graph retrieval: $q$, $\mathcal{G}$, $\mathbf{v}_q$ and $\mathcal{V}$.
    (b)--(d) depict representative graph retrieval methods, while (e) presents our \textit{GraFine} method.}
    \caption{
    Conceptual comparison of graph retrieval workflows based on retrieval operations.
    (a) illustrates the inputs for graph retrieval: $q$, $\mathcal{G}$, $\mathbf{v}_q$ and $\mathcal{V}$.
    (b)--(d) depict representative graph retrieval methods, while (e) presents our \textit{GraFine} method.
    }
    \label{fig:example}
\vspace{-0.3cm}    
\end{figure*}

Recently, increasing attention has been paid to \textit{corpus graphs}—such as reference networks~\cite{an-etal-2025-grex}—in which each node contains rich textual information~\cite{macavaney2022adaptive, rathee2025guiding, guo-etal-2025-lightrag, chen2025pathrag}. 
Unlike conventional knowledge graphs (KGs), nodes in corpus graphs generally encapsulate rich contextual clues that can guide the retrieval process. 
Accordingly, retrievers on corpus graphs can use such clues at \textit{retrieval time} to refine the retrieved set to move toward relevant yet initially unretrieved regions of the graph, thereby improving Graph RAG performance.
For example, in Figure~\ref{fig:example}, given a query about \textit{Agentic RAG components}, once the retriever has retrieved $n_3$ \textit{(Agentic RAG)}, it can use the corresponding content $c_3$ as a clue to infer that $n_4$ \textit{(ReAct)} is the next node to retrieve. 
This raises the following research question:

\begin{center}
\textit{How can we efficiently refine the intermediate retrieved set to reach previously unexplored but relevant regions on a corpus graph?}
\end{center}

Fundamentally, retrieval refinement can be described as an iterative process consisting of two steps: adding new elements to the current set of retrieved nodes (the \textit{adding} step) and pruning existing ones from it (the \textit{pruning} step).

Conventional approaches to address this fall into two broad categories. 
One line of work performs these two steps through retrieval-generation interleaving, including Agentic RAG methods that rely on repeated LLM reasoning~\cite{trivedi2023interleaving, lee-etal-2024-planrag, sun2024thinkongraph, ma2025thinkongraph}. 
However, such methods incur high latency, often reaching up to tens of seconds~\cite{shen2024tradeoffsretrievalaugmented, cook2025retrieval}. 
In interactive settings, such delays can substantially reduce usability~\cite{brutlag2008user, kim2025seconds} and hinder practical deployment in enterprise environments.

A second line of work instead combines relatively fast model-based \textit{pruning} (e.g., rerankers~\cite{macavaney2022adaptive}) with graph-search-based \textit{adding} methods, such as one-hop expansion~\cite{guo2025lightrag}, personalized PageRank~\cite{jimenez2024hipporag}, or PCST-based construction~\cite{he2024g}. 
While these methods are effective on many datasets, they fundamentally have two inherent \textit{informational} blind points: (1) \textbf{model-based pruning is topologically blind}, and (2) \textbf{graph search is semantically blind} in the current retriever design.
We later formalize this observation through \textit{an operational taxonomy} of graph retrieval methods, using it to clarify these blind points.
Together, these observations suggest two requirements for \textit{retrieval-time refinement} on corpus graphs: \textbf{topology-aware pruning} and \textbf{semantics-aware adding}, while remaining \textbf{time-efficient}.

To address this, we \textbf{expand the design of retrievers} and propose \textbf{GraFine}, a fast and effective realization of the proposed design.
After a seed retrieval step, GraFine alternates two lightweight refinement stages. 
The first is \textbf{Semantic Proximity eXpansion (SPX)}, which uses semantic signals to guide which nodes are added during expansion rather than relying on topology alone. 
The second is the \textbf{Graph Smoothing Reranker (GSR)}, which incorporates graph context during pruning and reranking rather than judging nodes from text alone. 
By interleaving these two stages, GraFine improves \textit{retrieval-time refinement} without incurring the latency of retrieval-generation interleaving. 

To evaluate the effectiveness of GraFine, we conduct both retrieval and generation experiments across two types of corpus graphs: \textbf{reference networks}~\cite{rohatgi-etal-2023-acl, an-etal-2025-grex, litfm_kdd25} and \textbf{text-rich knowledge graphs}~\cite{edge2024local, guo2025lightrag, chen2024examination, chen2025pathrag}. 
Across these settings, GraFine improves retrieval accuracy and downstream generation quality while remaining time-efficient, achieving an average gain of 9.9\% in Recall@10, generation win rates above 55\%, and a consistently better effectiveness--latency trade-off.

To further examine whether retrieval refinement moves the retrieved set closer to relevant but initially missed targets, we introduce a new retrieval metric, \textbf{Topological Recall}. Unlike standard Recall, which only measures whether oracle nodes are retrieved, Topological Recall additionally quantifies the \textit{topological proximity} between retrieved nodes and oracle nodes. 
Using this metric, we find that GraFine progressively moves intermediate retrieval results closer to oracle nodes, and that this progress is accompanied by improved recall.
Our main contributions are summarized as follows:
\begin{itemize}
    \item We formulate fast retrieval-time refinement on corpus graphs as \textbf{semantics-aware adding} and \textbf{topology-aware pruning} under efficiency constraints, and unify existing graph retrieval methods through an operational taxonomy.
    \item We propose \textbf{GraFine}, a lightweight retriever that alternates \textbf{SPX} for semantics-guided expansion and \textbf{GSR} for graph-aware pruning.
    \item We show on reference networks and text-rich knowledge graphs that GraFine improves retrieval and generation efficiently, and introduce \textbf{Topological Recall} to measure progress toward unretrieved oracle nodes.
\end{itemize}

%% file: sec2_preliminaries.tex
\section{Background and Problem Formulation}
\label{sec:preliminaries}

\subsection{Graph RAG on Corpus Graphs}
\label{subsec:corpus-graph}

In this paper, we focus on \textit{corpus graphs}, defined as follows:

\begin{definition}[Corpus Graph]
A graph $\mathcal{G}=(\mathcal{N}, \mathcal{E})$ is classified as a \textit{corpus graph} if and only if every node $n \in \mathcal{N}$ is associated with a pair $(k_n,c_n)$: $k_n$ serves as a node identifier (key), and $c_n$ provides descriptive textual information (textual content) about the node.
\end{definition}

Graph RAG methods for \textit{corpus graphs} take a query $q$ and a corpus graph $\mathcal{G}=(\mathcal{N}, \mathcal{E})$ as inputs, and generate an answer $a$ through the following two steps:

\begin{enumerate}
    \item \textbf{Graph Retrieval Step:} Given $q$ and $\mathcal{G}=(\mathcal{N}, \mathcal{E})$, the objective is to retrieve a set of nodes $\mathcal{N}_{\text{ret}} \subseteq \mathcal{N}$ that are \textit{relevant} to $q$.
    \item \textbf{Generation Step:} An answer $a$ is generated based on $q$ and the retrieved node contents $\mathcal{C}_{\mathrm{ret}}=\{c_i \mid (k_i, c_i) \in \mathcal{N}_{\mathrm{ret}}\}$ using a generative language model $P_{\theta}$. Conventionally, $P_{\theta}$ processes $q$ and $\mathcal{C}_{\mathrm{ret}}$ via in-context learning.
\end{enumerate}

In this study, we assume corpus graphs that may contain edge directions and cycles; and do not explicitly leverage edge-level textual information. 
Other graph types (e.g., graphs that contains textual descriptions on edges) are outside the scope of this work.


\subsection{Retrieval-Time Refinement}
\label{subsec:retrieval-time-refinement}

In corpus graphs, intermediate nodes encapsulate rich textual clues vital for moving toward structurally distant oracle nodes. 
We formalize this process of leveraging intermediate content to iteratively refine the retrieved node set as follows: 

\begin{definition}[Retrieval-time refinement]
Given a query $q$, a corpus graph $\mathcal{G}=(\mathcal{N},\mathcal{E})$, a set of initial seed nodes $\mathcal{N}_{\mathrm{seed}}=\mathcal{N}^{(0)} \subseteq \mathcal{N}$, and a set of oracle nodes $\mathcal{N}_q^\star \subseteq \mathcal{N}$ for $q$, \textit{retrieval-time refinement} is the iterative process of transforming the current node set $\mathcal{N}^{(t)}$ into a refined node set $\mathcal{N}_{\mathrm{ret}}=\mathcal{N}^{(T)}$ through a sequence of query-conditioned update steps:
\[
\mathcal{N}^{(t+1)} = \phi_t(q, \mathcal{N}^{(t)}, \mathcal{G}), \qquad t=0,\ldots,T-1.
\]
\end{definition}
Specifically, the update function $\phi_t$ performs two fundamental operations to approach the oracle nodes: (1) the \textit{adding} step, which incorporates new nodes $n^{+} \in \mathcal{N}^{(t+1)} \setminus \mathcal{N}^{(t)}$, and (2) the \textit{pruning} step, which filters out existing nodes $n^{-} \in \mathcal{N}^{(t)} \setminus \mathcal{N}^{(t+1)}$ from the current node set.

The objective is to transform $\mathcal{N}_{\mathrm{seed}}$ into a final set of nodes $\mathcal{N}_{\mathrm{ret}}$ that maximizes the inclusion of previously missed oracle nodes in $\mathcal{N}_q^\star$. 
For example, to successfully retrieve the oracle node $n_4$ for query $q$ in Figure~\ref{fig:example}(a) about \textit{components} of Agentic RAG, the retrieval must look beyond the isolated target content $c_4$ or its vector $\mathcal{V}_4$ and instead leverage the vital intermediate content $c_3$ of node $n_3$, which indicates that \textit{Agentic RAG employs multi-step prompting strategies such as ReAct}.



%% file: sec3_taxonomy.tex
\section{Rethinking Graph Retrieval: An Operational Taxonomy}
\label{sec:taxonomy}

\subsection{Fundamental Retrieval Operators}
\label{subsec:graph-retrieval-operators}

In a corpus graph $\mathcal{G}=(\mathcal{N}, \mathcal{E})$, a \textit{retrieval operator} $\mathcal{O}$ is defined as a composite function $\mathcal{P} \circ \mathcal{R}$, where a \textbf{ranking function} $\mathcal{R}$ scores nodes and a \textbf{pruning function} $\mathcal{P}$ selects the final subset $\mathcal{N}_{\mathrm{ret}}$.
To instantiate the retrieval-time refinement process defined in Section~\ref{subsec:retrieval-time-refinement}, we classify retrieval operators into three categories based on input sources and scoring mechanisms: Vector Search ($\vs$), Graph Search ($\gs$), and Model-based Search ($\model$). Specifically, $\vs$ initializes seed nodes, allowing $\gs$ and $\model$ to execute actual retrieval-time refinement via adding and pruning, respectively.
We detail their definitions below.

\subsubsection{Vector Search Operator}
The \textbf{Vector Search Operator}~($\vs$) serves as the seed initialization step. Instead of operating within the retrieval-time refinement loop, $\vs$ is responsible for generating the initial seed nodes that subsequent adding and pruning steps refine. 
Taking a query vector $\mathbf{v}_q$ and the set of all nodes $\mathcal{N}$ (associated with pre-indexed vectors $\mathcal{V}$) as inputs, it employs a ranking function $\mathcal{R}_\text{V}$ to compute vector similarity. This is followed by a pruning function $\mathcal{P}_\text{V}$ that identifies the top-$k$ nodes.
Formally, $\vs$ is defined as follows:

\begin{definition}[Vector Search, $\vs$]
\label{def:vs}
The Vector Search operator retrieves a node set $\mathcal{N}_{\mathrm{VS}}$ by
\begin{align*}
\mathcal{N}_{\text{V}} &= \vs(\mathbf{v}_q, \mathcal{N}, k) = \mathcal{P}_{\text{V}}(\{\mathcal{R}_{\text{V}}(\mathbf{v}_q, n)\mid n\in \mathcal{N}\}, k)\\
&= \operatorname*{arg\,topk}_{n \in \mathcal{N}} \; \operatorname{sim}(\mathbf{v}_q, \mathbf{v}_n)
\end{align*}
where $\mathbf{v}_n$ denotes the vector representation of node $n$, and $\operatorname{sim}(\cdot, \cdot)$ denotes a vector similarity metric.
\end{definition}

\begin{example}[Dense Vector Search]
\label{eg:dense-vector-search}
In dense retrieval, the node vectors lie in a continuous latent manifold $\mathbb{R}^d$. The similarity is typically defined as the cosine similarity:
\[
    \operatorname{sim}(\mathbf{v}_q, \mathbf{v}_n) = \frac{\mathbf{v}_q^\top \mathbf{v}_n}{|\mathbf{v}_q| \cdot |\mathbf{v}_n|}
\]
\end{example}

\subsubsection{Graph Search Operator}


The \textbf{Graph Search Operator}~($\gs$) functions as the \textit{adding} step during refinement in retrieval-time. It takes a set of seed nodes $\mathcal{N}_{\text{seed}} \subset \mathcal{N}$, seed node features $\mathbf{H}_{\text{seed}}$, and the edge set $\mathcal{E}$ as inputs.
It employs a ranking function $\mathcal{R}_{\text{G}}$ to calculate topological scores $\mathbf{H}^{(L)}$ via $L$-step signal propagation on $\mathcal{E}$, generalizing Graph Neural Networks (GNNs) message-passing. 
By depending on $\mathcal{N}_\text{seed}$ and $\mathbf{H}_{\text{seed}}$, $\gs$ to perform expansion based on the intermediate retrieved set, making it the primary mechanism for \textit{adding} new nodes during iterative refinement. We define $\gs$ as follows:

\begin{definition}[Graph Search, $\gs$]
\label{def:gs}
The Graph Search operator retrieves a node set $\mathcal{N}_{\mathrm{G}}$ by 
\begin{align*}
    \mathcal{N}_{\mathrm{G}} &= \gs(\mathcal{N}_{\text{seed}}, \mathbf{H}_{\text{seed}}, \mathcal{E}) =  \mathcal{P}_{\mathrm{G}}(\mathcal{R}_{\mathrm{G}}(\mathcal{N}_{\text{seed}}, \mathbf{H}_{\text{seed}}, \mathcal{E}))\\
    & =\mathcal{P}_{\mathrm{G}}\left( \mathbf{H}^{(L)}\right)
\end{align*}
Here, $\mathbf{H}^{(L)}$ denotes the node scores in $\mathcal{N}$ after $L$ steps. 
With $\mathbf{H}^{(0)}=\mathbf{H}_\text{seed}$ initialized by the preceding retrieval, the $l$-th update rule is:
\begin{equation*}
    \mathbf{H}^{(l)} = f_{\text{prop}}^{(l)} \left(\mathcal{N}_{\text{seed}}, \mathbf{H}^{(l-1)}, \mathcal{E}\right) 
\end{equation*}
\end{definition}

\begin{example}[1-hop Traversal]
In 1-hop traversal-based retrieval~\cite{guo2025lightrag}, $\mathcal{R}_{\mathrm{G}}$ is a single-layer propagation ($L=1$) where $f_{\text{prop}}^{(1)}$ is defined as:
\[
    \mathbf{H}^{(1)} = f_{\text{prop}}^{(1)}(\mathcal{N}_{\text{seed}}, \mathbf{H}^{(0)}, \mathcal{E}) = \mathbf{A}\mathbf{H}^{(0)}
\]
where $\mathbf{A}$ is the adjacency matrix derived from $\mathcal{E}$. Here, $\mathbf{H}^{(0)}$ is initialized such that $[\mathbf{H}^{(0)}]_n = \mathbb{I}(n\in \mathcal{N}_{\text{seed}})$.
The pruning function $\mathcal{P}_{\mathrm{G}}$ selects immediate neighbors by excluding the seed nodes themselves:
\[
    \mathcal{P}_{\mathrm{G}}(\mathbf{H}^{(1)}) = \{ n \mid [\mathbf{H}^{(1)}]_n > 0 \} \setminus \mathcal{N}_{\text{seed}}
\]
\end{example}

\begin{example}[PageRank Retrieval]
\label{eg:ppr}
In PageRank-based retrieval~\cite{jimenez2024hipporag, gutierrez2025from}, $\mathcal{R}_{\mathrm{G}}$ performs iterative propagation using the update function $f_{\text{prop}}^{(l)}$ until $\mathbf{H}^{(L)}$ converges, where $f_{\text{prop}}^{(l)}$ is defined as:
\[
    \mathbf{H}^{(l)} = f_{\text{prop}}^{(l)}(\mathcal{N}_{\text{seed}}, \mathbf{H}^{(l-1)}, \mathcal{E}) = (1-\alpha)\mathbf{M}\mathbf{H}^{(l-1)} + \alpha\mathbf{H}^{(0)}
\]
where $\alpha$ denotes the restart probability and $\mathbf{M}$ is the column-normalized transition matrix derived from $\mathcal{E}$. Here, $\mathbf{H}^{(0)}$ serves as the personalized restart distribution.
The pruning function $\mathcal{P}_{\mathrm{G}}$ selects the top-$k$ nodes with the highest scores as follows:
\[
    \mathcal{P}_{\mathrm{G}}(\mathbf{H}^{(L)}) = \operatorname*{arg\,topk}_{n \in \mathcal{N} \setminus \mathcal{N}_\text{seed}} [\mathbf{H}^{(L)}]_n
\]
\end{example}

\subsubsection{Model-based Search}

Third, the \textbf{Model-based Search ($\model$)} operator serves as the \textit{pruning} step within the refinement process. While $\gs$ refines the set by adding candidate nodes, $\model$ evaluates and \textit{prunes} this intermediate retrieved set to retain only highly relevant nodes. It takes a textual query $q$ and a set of seed nodes $\mathcal{N}_{\mathrm{seed}} \subseteq \mathcal{N}$ as inputs.
In its ranking function $\mathcal{R}_{\mathrm{M}}$, it utilizes a computationally intensive model $P_\phi$ (e.g., a language model) to process the raw textual node contents and assess their semantic relevance.
By enabling late query--content interaction, $\model$ can identify and downrank unhelpful nodes that are semantically irrelevant.
Due to the high computational cost of this interaction, the operator is typically restricted to evaluating the smaller subset $\mathcal{N}_{\mathrm{seed}} \subset \mathcal{N}$ generated by the preceding steps.
Formally, $\model$ is defined as follows:

\begin{definition}[Model-based Search, $\model$]
\label{def:model}
The Model-based Search operator selects a node set $\mathcal{N}_{\mathrm{M}}$ from $\mathcal{N}_\text{seed} \subset \mathcal{N}$ by
\begin{align*}
\mathcal{N}_{\text{M}} &= \model(q, \mathcal{N}_\text{seed}, k) = \mathcal{P}_{\text{M}}(\{\mathcal{R}_{\text{M}}(q, n)\mid n\in \mathcal{N}_\text{seed}\}, k)\\
&= \operatorname*{arg\,topk}_{n \in \mathcal{N}_\text{seed}} \; P_\phi(q, n)
\end{align*}
\end{definition}

\subsection{Unifying Existing Graph Retrieval Methods}
\label{subsec:existing-methods}

The three retrieval operators defined in Section~\ref{subsec:graph-retrieval-operators} serve as the foundational units shared by the representative retrieval methods. 
As shown in Table~\ref{tab:graph-rag-comparison}, existing methods that appear algorithmically diverse can be expressed as different compositions of $\vs, \gs$ and $\model$.

For example, a standard Retrieve-then-Rerank pipeline (Example~\ref{eg:re2}) is formally expressed as $\vs \rightarrow \model$: vector search first produces a candidate set, which is subsequently pruned by a model-based reranker.
Similarly, G-Retriever (Example~\ref{eg:g-retriever}) is expressed as $\vs \rightarrow \gs$, where initial seed nodes are identified via vector search and then expanded into the final retrieved set via graph-based search. 
Our proposed taxonomy provides a unified lens that encompasses a broad spectrum of representative graph retrieval methods, such as LightRAG~\cite{guo2025lightrag}, PathRAG~\cite{chen2025pathrag}, and GAR~\cite{macavaney2022adaptive} shown in Figure~\ref{fig:example}(b--d). Ultimately, the decomposition of existing methods in Table~\ref{tab:graph-rag-comparison} demonstrates that \textit{seemingly disparate methods can be rewritten using our fundamental operator taxonomy}.


\begin{example}[Retrieve-then-Rerank Pipeline]
\label{eg:re2}
Consider a standard pipeline employing a Bi-encoder (e.g., Contriever) for retrieval and a Cross-encoder (e.g., BERT) for reranking. While the overall process is neither $\vs$ nor $\model$, we can divide the pipeline into two subprocesses:

\begin{enumerate}
    \item Candidate Retrieval: The Bi-encoder retrieves the top-$100$ candidates ($\mathcal{N}_{\text{cand}}$) based on cosine vector similarity with query vector $\mathbf{v}_q=\text{Contriever}(q)$. This instantiates $\vs$:
    \[
        \mathcal{N}_{\text{cand}} = \operatorname*{arg\,top100}_{n \in \mathcal{N}} \; \frac{\mathbf{v}_q^\top \mathbf{v}_n}{|\mathbf{v}_q| \cdot |\mathbf{v}_n|}= \vs(\mathbf{v}_q, \mathcal{N}, 100)
    \]
    
    \item Reranking: The Cross-encoder scores $\mathcal{N}_{\text{cand}}$ given the query $q$ to select the final top-$10$ nodes. This instantiates $\model$:
    \[
        \mathcal{N}_{\text{final}} = \operatorname*{arg\,topk}_{n \in \mathcal{N}_\text{seed}} \; MLP(BERT(q \oplus n)) = \model(q, \mathcal{N}_{\text{cand}}, 10)
    \]
\end{enumerate}
\end{example}

\begin{example}[G-Retriever]
\label{eg:g-retriever}
The retrieval in G-Retriever~\cite{he2024g} comprises two stages: (1) Vector-Edge Retrieval ($\vs$) and (2) PCST-based subgraph construction ($\gs$), as detailed below:
\begin{enumerate}
\item Vector-Edge Retrieval ($\vs$): It retrieves the top-$k$ $\mathcal{N}_\text{sub}\subset\mathcal{N}$ and $\mathcal{E}_\text{sub}\subset\mathcal{E}$ via vector similarity.
For the node and edge at rank $i$, we assign rank-based prizes $k-i$ to initialize $\mathbf{H}_{\text{seed}}$.
\item PCST Construction ($\gs$): The operator selects the final subgraph $\mathcal{N}_{\text{ret}}$ and $\mathcal{E}_{\text{ret}}$ by optimizing the PCST ranking function: 
\[
\mathcal{R}_{\text{PCST}}(\mathcal{N}_{\text{ret}}, \mathcal{E}_{\text{ret}}) = \sum_{n \in \mathcal{N}_{\text{ret}}} p(n) + \sum_{e \in \mathcal{E}_{\text{ret}}} p(e) - c(\mathcal{N}_{\text{ret}}, \mathcal{E}_{\text{ret}})
\]
Here, $p(n)$ and $p(e)$ correspond to the values in $\mathbf{H}_{\text{seed}}$, and $c(\mathcal{N}_{\text{ret}}, \mathcal{E}_{\text{ret}})$ only depends on $\mathcal{E}$.
\end{enumerate}
\end{example}

\begin{table}[t!]
\centering
\caption{Comparison of representative Graph RAG methods based on target database and retrieval operators.}
\vspace{-0.2cm}
\renewcommand{\arraystretch}{0.9}
\resizebox{\columnwidth}{!}
{
\begin{tabular}{l l c c c c c} 
\toprule
\multirow{2}{*}{\textbf{Target database}} & \multirow{2}{*}{\textbf{Methods}} & \multicolumn{3}{c}{\textbf{Basic operations}} & \multicolumn{2}{c}{\textbf{Fusion operators}} \\ 
\cmidrule(lr){3-5} \cmidrule(lr){6-7}
 & & \textbf{\vs} & \textbf{\gs} & \model & \textbf{\vgs} & \textbf{\gmodel} \\ \midrule
 & HyKGE~\cite{jiang2024hykgehypothesisknowledgegraph} & \cmark & \xmark & \xmark & \xmark & \xmark \\
\cellcolor{white} & HippoRAG~\cite{jimenez2024hipporag} & & & & & \\
\cellcolor{white} & GNN-RAG~\cite{mavromatis2024gnn} & & & & & \\
\cellcolor{white} & G-Retriever~\cite{he2024g} & \multirow{-3}{*}{\cmark} & \multirow{-3}{*}{\cmark} & \multirow{-3}{*}{\xmark} & \multirow{-3}{*}{\xmark} & \multirow{-3}{*}{\xmark} \\
 & SubgraphRAG~\cite{li2025simple} & \cmark & \xmark & \cmark & \xmark & \xmark \\
\cellcolor{white} & LightPROF~\cite{ao2025lightprof} & \cmark & \cmark & \cmark & \xmark & \xmark \\
\multirow{-7}{*}{Knowledge graph} & ToG~\cite{sun2024thinkongraph} & \cmark & \cmark & \cmark & \xmark & \xmark \\ \cmidrule(lr){1-7}
\cellcolor{white} & LightRAG~\cite{guo2025lightrag} & & & & & \\
 
\cellcolor{white} & PathRAG~\cite{chen2025pathrag} & \multirow{-2}{*}{\cmark} & \multirow{-2}{*}{\cmark} & \multirow{-2}{*}{\xmark} & \multirow{-2}{*}{\xmark} & \multirow{-2}{*}{\xmark} \\ 

\cellcolor{white}\multirow{-3}{*}{Corpus graph} & GRAG~\cite{hu-etal-2025-grag} & \cmark & \xmark & \xmark & \xmark & \xmark \\
\cmidrule(lr){1-7}
\cellcolor{white}{Corpus graph + KG} & KG2RAG~\cite{zhu-etal-2025-knowledge} & \cmark & \cmark & \cmark & \xmark & \xmark \\\cmidrule(lr){1-7}
\cellcolor{white} & ToG 2.0~\cite{ma2025thinkongraph} & \cmark & \cmark & \cmark & \xmark & \xmark \\ 
\cellcolor{white} \multirow{-2}{*}{Documents + KG}  & HippoRAG 2~\cite{gutierrez2025from} & \cmark & \cmark & \xmark & \xmark & \xmark \\ 

\cmidrule(lr){1-7} 
\cellcolor{white} \multirow{-1}{*}{Corpus graph}& \textbf{GraFine (ours)} & \cmark & - & - & \cmark & \cmark \\ 
\bottomrule
\end{tabular}
}
\label{tab:graph-rag-comparison}
\end{table}

More importantly, this decomposition exposes the \textbf{characteristic blind spots} that limit retrieval-time refinement in current methods. Since each operator can only rank nodes based on the information in its inputs, current operators are inherently restricted in how it performs the crucial \textit{adding} and \textit{pruning} of nodes.

\begin{itemize}
    \item \textbf{$\gs$ performs semantically-blind \textit{adding}}: While $\gs$ effectively propagates signals over graph topology, it lacks direct query--content interaction when identifying new candidate nodes. Thus, during the \textit{adding} step, $\gs$
    often drifts toward topologically reachable but semantically irrelevant nodes (e.g., node $n_5$ representing \textit{Hallucination} in Figure~\ref{fig:example}(d)), failing to expand in a contextually relevant direction.
    \item \textbf{$\model$ performs topologically-blind \textit{pruning}}: Conversely, while $\model$ enables rich query--content interaction, it evaluates nodes in isolation from the graph structure. 
    \textit{Pruning} through $\model$ relies solely on local semantic similarity, causing it to assign misleading low scores to crucial nodes (e.g., node $n_4$ in Figure~\ref{fig:example}(d)) whose relevance is only established in the structural links and contextual clues provided by surrounding intermediate nodes. 
\end{itemize}

Consequently, existing methods composed of these foundational operators inherit these blind spots as systemic limitations. 
Methods relying on $\gs$ for adding are prone to drifting toward \textit{topologically reachable but semantically unhelpful} nodes, while methods using $\model$ for pruning often fail to retain nodes whose relevance emerges \textit{only through graph context}. Since any composition of these operators inherently suffers from these blind spots, current methods are restricted in their ability to perform effective retrieval-time refinement.



%% file: sec3_methodologies.tex
\vspace{-0.1cm}
\section{GraFine: Retrieval-Time Refinement for Graph RAG}
\label{sec:methodology}

\subsection{Design: Semantic–Topological Refinement}
\label{subsec:fusion-operators}

To achieve effective retrieval-time refinement and move beyond the systemic limitations of existing methods, it is important to resolve the blind spots inherent in the fundamental operators.
To this end, we propose two novel fusion operators which leverage both semantic and topological inputs, ensuring the \textit{adding} and \textit{pruning} steps are semantically guided and topologically aware, respectively. 

First, to overcome the \textit{semantic blindness} inherent in $\gs$, we define a \textbf{Vector-Graph Search} ($\vgs$) operator. Rather than drifting toward topologically reachable but \textit{unhelpful} nodes, $\vgs$ expands the retrieved set by utilizing both the query vector $\mathbf{v}_q$ and graph structure $\mathcal{E}$. This dual integration ensures that the addition of new nodes is steered toward semantically valuable regions.
\begin{definition} [Vector-Graph Search, $\vgs$]
\label{def:vgs}
    $\vgs$ retrieves nodes using both vector representations $\mathcal{V}$ and graph topology $\mathcal{E}$:
    \begin{equation*}
        \mathcal{N}_{\text{VG}} = \vgs(\mathbf{v}_q, \mathcal{V}, \mathcal{N}_\text{seed}, \mathcal{E})
    \end{equation*}
\end{definition}

Second, to overcome the \textit{topological blindness} inherent in $\model$, we define a \textbf{Graph-Model Search} ($\gmodel$) operator. Instead of prematurely deleting nodes that lack standalone semantic similarity, $\gmodel$ integrates the structural context of surrounding nodes $\mathcal{E}$ directly into relevance scoring. This ensures the pruning process safely retains nodes whose true value emerges only through the structural context. 
\begin{definition}[Graph-Model Search, $\gmodel$]
\label{def:gmodel}
$\gmodel$ selects nodes from $\mathcal{N}_\text{seed}\subset \mathcal{N}$ by incorporating topological context $\mathcal{E}$:
\begin{align*}
\mathcal{N}_{\text{GM}} &= \gmodel(q, \mathcal{N}_\text{seed}, \mathcal{E}, k) = \mathcal{P}_{\text{GM}}(\mathcal{R}_{\text{GM}}(q, \mathcal{N}_\text{seed}, \mathcal{E}), k)
\end{align*}
\end{definition}

\subsection{GraFine Framework}
\label{subsec:GraFine}

The \textbf{GraFine} framework is the practical realization of our proposed retriever design, developed to execute effective retrieval-time refinement by resolving blind spots. 
GraFine takes the following inputs: a query $q$, a corpus graph $\mathcal{G}=(\mathcal{N}, \mathcal{E})$, a query vector $\mathbf{v}_{q}$, a set of node vectors $\mathcal{V}$, and a set of \textit{hyperparameters} ($BATCH, \alpha, \beta, \text{and } b_{\max}$). The output is a final, refined set of retrieved nodes, $\mathcal{N}_\text{ret}$.

The execution flow of GraFine is outlined in Algorithm~\ref{alg:GraFine}.
In this algorithm, each colored box represents a retrieval operator: $\colorbox{pastelgreen}{\vs}$ (Vector Search), $\colorbox{pastelblue}{\gmodel}$ (Graph Model-based Search), and $\colorbox{pastelpurple}{\vgs}$ (Vector-Graph Search).
The process comprises two primary phases: the \textbf{initial setup step} and the \textbf{iterative retrieval step}. We explain these steps in detail below.
Here, SPX and GSR stand for Semantics-Guided Graph Expansion~(Section~\ref{subsec:SPX}) and Graph-Aware Reranking~(Section~\ref{subsec:GSR}), respectively, which are proposed in this paper.

\begin{algorithm}[!htb]
\caption{The GraFine framework}
\label{alg:GraFine}
\begin{algorithmic}[1]
\Require Query $q$, Graph $\mathcal{G}=(\mathcal{N}, \mathcal{E})$, Node vectors $\mathcal{V} = \{Enc_{N}(n_i) \mid n_i \in \mathcal{N}\}$, Query vector $\mathbf{v}_{q}=Enc_{Q}(q)$, Batch size $BATCH$, Smoothing factor $\alpha$, Score ratio $\beta$, Budget $b_{\max}$
\Ensure $\mathcal{N}_\text{ret}$ (Set of retrieved nodes)


\State \colorbox{pastelgreen}{$\mathcal{N}_\text{ret} \gets \operatorname*{arg\,top}{BATCH}_{n \in \mathcal{N}} \; \operatorname{sim}(\mathbf{v}_q, \mathcal{V}_n)$}  \Comment{1. Initial $\vs$}

\State \colorbox{pastelblue}{$\mathcal{N}_\text{ret} \gets \textbf{GSR}(q,\mathcal{N}_\text{ret}, \mathcal{E}, \alpha)$} \Comment{2. $\gmodel$ using \textit{GSR}}

\While{$|\mathcal{N}_\text{ret}| < b_{\max}$} \Comment{3. Expansion Loop with $\vgs$}
    
    \State \colorbox{pastelpurple}{$\mathcal{N}_\text{add} \gets \mathbf{SPX}( \mathbf{v}_q, \mathcal{V}, \mathcal{E}, \mathcal{N}_\text{ret}, \beta)$} \Comment{$\vgs$ using \textit{SPX}}
    
    \State $k_{remain} \gets \min(|\mathcal{N}_\text{ret}| + BATCH, b_{\max}) - |\mathcal{N}_\text{ret}|$
    \State $\mathcal{N}_\text{ret} \gets \mathcal{N}_\text{ret} \cup \mathcal{N}_\text{add}[:k_{remain}]$ \Comment{Apply hard budget cap}
    
    \State \colorbox{pastelblue}{$\mathcal{N}_\text{ret} \gets \textbf{GSR}(q,\mathcal{N}_\text{ret}, \mathcal{E}, \alpha)$} \Comment{$\gmodel$ using \textit{GSR}} 
\EndWhile

\State \Return $\mathcal{N}_\text{ret}$
\end{algorithmic}
\end{algorithm}

\noindent \textit{Initial setup step (Lines 1--2).} In this step, the retriever establishes the starting nodes for the subsequent iterative process.
\begin{itemize}
    \item \textbf{(L1)} First, a dot product-based vector search is performed on the node vectors $\mathcal{V}$ to retrieve a top-$BATCH$ list of candidates, $\mathcal{N}_\text{ret}$, sorted by their dot product scores.
    \item \textbf{(L2)} The \textbf{GSR} method is applied to assign initial relevance scores to these nodes.
\end{itemize}

By employing this iterative process~\textit{(Lines 4--7)}, GraFine enhances the quality of the input nodes $\mathcal{N}_\text{ret}$ fed into GSR, allowing us to produce better retrieval outcomes while minimizing the usage of GSR (i.e., minimize computational cost).
\begin{itemize}
    \item \textbf{(L4)} In each iteration, our $\mathbf{SPX}$ algorithm~($\vgs$ operator)~identifies potential new nodes~($\mathcal{N}_\text{add}$)~based on the current $\mathcal{N}_\text{ret}$.
    \item \textbf{(L5--L6)} The retriever then incorporates up to $BATCH$ new nodes into $\mathcal{N}_\text{ret}$ (strictly adhering to the budget cap $b_{\max}$).
    \item \textbf{(L7)} For the next iteration, the retriever re-applies \textbf{GSR} to update the rankings of the retrieved nodes $\mathcal{N}_\text{ret}$.
\end{itemize}

The hyperparameters control the algorithm's behavior as follows: $BATCH$ denotes the number of nodes added to $\mathcal{N}_\text{ret}$ during a single iteration; $\alpha$ represents the smoothing factor for GSR (detailed in Section~\ref{subsec:GSR}); $\beta$ represents the score ratio for SPX (in Section~\ref{subsec:SPX}) and $b_{\max}$ defines the maximum node budget, serving as the stopping criterion for the iterative loop.

\subsection{SPX: Semantics-Guided Graph Expansion}
\label{subsec:SPX}

We propose \textbf{Semantic Proximity eXpansion (SPX)}, a lightweight implementation of the $\vgs$ that identifies candidates to \textit{add} by leveraging both topological structure and semantic representations, unlike conventional topology-only methods. 
As detailed in Algorithm~\ref{alg:SPX}, the procedure ranks candidates in $\mathcal{N}_\text{SPX}$ using a composite score—a $\beta$-weighted sum of structural importance ($I_{Struct}$) and semantic similarity ($I_{Sim}$):
\begin{itemize}
\item \textbf{$I_{Struct}$ (Lines 3--12):} This score combines \textit{rank proximity} and \textit{bridging capability}. 
It captures proximity to high-ranking context by favoring candidates connected to the highest-ranked nodes ($r_{best}$) in $\mathcal{N}_\text{ret}$. Additionally, it includes a bridging factor $|A(n)|$ that favors nodes linked to multiple retrieved nodes across the graph structure.
\item \textbf{$I_{Sim}$ (Line 13):} This is the dot product similarity between the query $\mathbf{v}_q$ and the candidate vector $\mathcal{V}_n$. This helps preserve semantically relevant nodes even if not topologically central.
\end{itemize}

\begin{algorithm}[t]
\caption{Our SPX method for $\vgs$}
\label{alg:SPX}
\begin{algorithmic}[1]
\Require Query vector $\mathbf{v}_q$, Node vectors $\mathcal{V}$, Edges $\mathcal{E}$, Retrieved $\mathcal{N}_\text{ret}$, Score ratio $\beta$. 
\InnerFunc $\mathrm{rankCheck}(n, \mathcal{N}_\text{ret})$ checks the rank of $n$, $\mathrm{deg}_{\mathcal{E}}(n)$ calculates the \textit{total} degree of the node $n$.
\Ensure Set of nodes to add $\mathcal{N}_\text{add}$

\State $\mathcal{N}_\text{SPX} \gets \{n_j \mid \exists_{n_i \in \mathcal{N}_\text{ret}} (n_i, n_j) \in \mathcal{E}\} \setminus \mathcal{N}_\text{ret}, R_{max} \gets |\mathcal{N}_\text{ret}|$
\For{$n \in \mathcal{N}_\text{SPX}$}
    \State $I_{Struct} \gets 0$
    \State $A(n) \gets \{v \in \mathcal{N}_\text{ret} \mid (v,n) \in \mathcal{E}\}$ \Comment{Adjacent retrieved nodes}
    \If{$R_{max} > 1$}
        \State $r_{best} \gets \min \{\mathrm{rankCheck}(v, \mathcal{N}_\text{ret}) \mid v \in A(n)\}$
        \State ${I_{Struct}} \gets 1 - \frac{r_{best} - 1}{R_{max} - 1}$ 
    \EndIf
    \State $C_{max} \gets \min(\mathrm{deg}_{\mathcal{E}}(n), R_{max})$
    \If{$C_{max} > 1$}
        \State $I_{Struct} \gets I_{Struct}+ \frac{|A(n)| - 1}{C_{max} - 1}$ \Comment{1. \textit{Structural score ($I_{Struct}$)}}
    \EndIf
    \State $I_{Sim} \gets \mathbf{v}_q \cdot \mathcal{V}_n$ \Comment{2. \textit{Similarity score} ($I_{Sim}$)}

    \State $\mathbf{S}_n \gets  I_{Sim} + \beta\cdot (I_{Struct})$
\EndFor

\State $\mathcal{N}_\text{add} \gets \texttt{argsort}(\mathcal{N}_\text{SPX}, \texttt{score}=\mathbf{S})$
\State \Return $\mathcal{N}_\text{add}$
\end{algorithmic}
\end{algorithm}

\subsection{GSR: Graph-Aware Reranking}
\label{subsec:GSR}

As a practical implementation of the $\gmodel$ operator defined in Definition~\ref{def:gmodel}, we propose the \textbf{Graph Smoothing Reranker (GSR)}.
To overcome the \textit{topological blindness} in conventional model search, we interpret the initial cross-encoder embeddings $\mathbf{H}$ as structurally incomplete, noisy signals, as they rely solely on isolated text semantics.
Consequently, GSR treats the task as a \textit{graph signal denoising problem}, aiming to smooth $\mathbf{H}$ by leveraging $\mathcal{E}$.
This corresponds to minimizing the Laplacian-regularized objective:
\[
\mathcal{L}(\mathbf{H}') = \frac{1}{2}\|\mathbf{H}'-\mathbf{H}\|_F^2 + \frac{\lambda}{2}\mathrm{Tr}(\mathbf{H}'^\top \mathbf{L}_{rw}\mathbf{H}')
\]
where $\mathbf{L}_{rw}=\mathbf{I}-\mathbf{P}$ is the random-walk Laplacian.
Instead of the computationally expensive closed-form solution, we employ a \textit{first-order approximation} via a single gradient descent step. This yields our efficient update rule:
\[
\mathbf{H}' \leftarrow \mathbf{H} - \eta \nabla \mathcal{L}(\mathbf{H}) = (1-\alpha)\mathbf{H}+\alpha(\mathbf{P}\mathbf{H})
\]
where $\alpha=\eta\lambda$.
Algorithm~\ref{alg:GSR} details this process, where Lines 2--7 introduce our modifications to the standard \textit{pruning} workflow.
The detailed procedure of these steps is as follows:

\begin{itemize}
\item \textbf{(L2--L5) Propagation Matrix Construction:} GSR constructs a normalized propagation matrix $\mathbf{P}$ from subgraph adjacency ($\mathbf{A}$) and degree ($\mathbf{D}$) matrices. Using reciprocal node degrees, $\mathbf{P}$ moderates high-degree node influence during aggregation.
\item \textbf{(L6) Latent Graph Fusion:} Initial latent vectors $\mathbf{H}$ are smoothed with neighbor-aggregated context ($\mathbf{P} \cdot \mathbf{H}$) via convolution. The factor $\alpha$ controls the trade-off between semantics and structural support, yielding fused representations $\mathbf{H}'$.
\item \textbf{(L7) Semantic Scoring:} The final relevance scores $\mathbf{S}$ are computed by passing $\mathbf{H}'$ through the MLP head, so the ranking reflects both semantic relevance and topological evidence.
\end{itemize}

\begin{algorithm}[t]
\caption{Our GSR method for $\gmodel$}
\label{alg:GSR}
\begin{algorithmic}[1]
\Require Query $q$, Retrieved $\mathcal{N}_\text{ret}$, Edges $\mathcal{E}$, Smoothing factor $\alpha$
\Ensure Reranked list of nodes $\mathcal{N}_\text{ret}$
\InnerFunc $\mathrm{Encoder}(\cdot)$ for latent vector extraction, $\mathrm{MLP}(\cdot)$ for scoring.

\State $\mathbf{H} \gets [\mathrm{Encoder}(q, n_i)]_{n_i \in \mathcal{N}_\text{ret}}$ \Comment{Extract Latent Vectors}
\State $\mathbf{A}\!\in\!\{0,1\}^{|\mathcal{N}_\text{ret}|\times|\mathcal{N}_\text{ret}|} \text{ where } \mathbf{A}_{ij}\!=\! \mathbb{I}((n_i, n_j) \in \mathcal{E})$ 
\State $\mathbf{D} \in \mathbb{R}^{|\mathcal{N}_\text{ret}|\times|\mathcal{N}_\text{ret}|} \text{ where } D_{ii} = \sum_j A_{ij}$ 
\State $\textbf{W} \gets \textbf{A} \cdot \textbf{D}^{-1}$ \Comment{Weighting by reciprocal of degrees}
\State $\textbf{P} \gets \text{diag}(\textbf{W} \cdot \mathbf{1})^{-1}\cdot \textbf{W}$ \Comment{Normalized Propagation Matrix}
\State $\mathbf{H}' \gets (1-\alpha)\mathbf{H} + \alpha (\textbf{P}\mathbf{H})$ \Comment{\textbf{Latent Graph Fusion}}
\State $\mathbf{S} \gets \mathrm{MLP}(\mathbf{H}')$ \Comment{Scoring via Classifier Head}
\State $\mathcal{N}_\text{ret} \gets \texttt{argsort}(\mathcal{N}_\text{ret}, \texttt{score=}\mathbf{S})$
\State \Return $\mathcal{N}_\text{ret}$
\end{algorithmic}
\end{algorithm}

\subsection{Efficiency Analysis}
\label{subsec:complexity-analysis}

In this section, we analyze the efficiency of GraFine with a focus on the two refinement components, SPX for $\vgs$ and GSR for $\gmodel$. 
Overall, after the initial retrieval, a query performs $T$ executions of SPX and $T+1$ executions of GSR, where $T$ is the number of refinement iterations until the budget cap is reached. We compare the practical efficiency of these components with related methods on real hardware in Section~\ref{subsubsec:query-processing-time}.

\vspace{1mm}
\noindent \textit{For $\vgs$.} SPX scores only the candidates reachable from the current retrieved nodes. Under an adjacency-list implementation with precomputed node degrees and a lookup table for the current ranks, each execution scans the relevant outgoing edges once to construct $N_{\text{add}}$, computes the structural scores, and then evaluates vector similarities for those candidates. 
Therefore, one SPX call is linear in the size of the explored nodes, up to the candidate-sorting step, which adds $O(|N_{\text{add}}|\log |N_{\text{add}}|)$. Accordingly, the total cost of $\vgs$ is the sum of these costs across all iterations, which is relatively small in our experimental observation.

\vspace{1mm}
\noindent \textit{For $\gmodel$.} Each GSR execution reranks only the current retrieved set. The dominant semantic computation remains the PLM-based encoding in Line 1 of Algorithm~\ref{alg:GSR}, but we cache the latent vectors $\mathbf{H}$, so each node is encoded only when it first enters $\mathcal{N}_\text{ret}$; hence, across the whole query, the total encoding cost grows with the number of distinct retrieved nodes, which is bounded by $b_{\max}$, rather than with the number of GSR calls. The additional graph operations incur a cost of $O(|\mathcal{N}_\text{ret}|^2)$ per execution, but they are applied only to the current retrieved set. 
We examine the practical effect of these local overheads in Section~\ref{subsubsec:query-processing-time}.

%% file: sec4_experiments.tex
\section{Experimental Evaluation}
\label{sec:experiments}

We conduct two types of experiments: (1) \textit{a retrieval experiment}, which aims to retrieve $\mathcal{N}_\text{ret}$ for a given query $q$, and (2) \textit{a RAG experiment}, which focuses on generating responses based on $\mathcal{N}_\text{ret}$. 
Unless otherwise specified, we use OpenAI's text-embedding-3-small as the embedding model, OpenAI's gpt-5-mini as the generative LLM, and bge-reranker-v2-m3 as the reranker.
For validating efficiency, we use two server configurations: (a) eight NVIDIA 24GB TITAN GPUs and (b) six NVIDIA 80GB A100 GPUs. 
We use these two hardware configurations to demonstrate that the efficiency trends of the compared methods remain robust across different system environments.
All other experiments, except those for efficiency evaluation, are conducted via configuration (a).

\subsection{Baselines}
\label{subsec:implement}

\subsubsection{Retrieval Baselines}
We evaluate retrieval performance using five document retrieval baselines and four graph retrieval baselines.
While many Graph RAG methods designed for KGs are incompatible with corpus graphs, we include HippoRAG 2~\cite{gutierrez2025from} by adapting it to the corpus graph setting.
Other methods that cannot be applied to corpus graphs are excluded.

Five document retrieval baselines rely on vector search ($\vs$) or model-based reranking ($\model$).
We include 
(1) \textbf{Vector Search} ($\vs$): retrieves nodes based on dot product similarity;
(2-3) \textbf{SPLADE}~\cite{formal2021splade} ($\vs$) and \textbf{Contriever}~\cite{izacard2022unsupervised} ($\vs$): representative sparse and dense retrieval methods, respectively;
(4) \textbf{HyDE}~\cite{gao-etal-2023-precise} ($\vs$): performs retrieval using generated hypothetical documents; and
(5) \textbf{Retrieve-then-Rerank (Re2)} ($\vs, \model$): reranks the top-100 candidates retrieved by Vector Search.

Four graph retrieval baselines and GraFine incorporate graph topology via the graph search operator~($\gs$).
We include
(1) \textbf{GAR} \cite{macavaney2022adaptive} ($\vs, \gs, \model$): dynamically interleaves vector and graph search with iterative reranking ($b_{\max}=100, BATCH=10$);
(2-3) \textbf{LightRAG/PathRAG}~\cite{guo2025lightrag, chen2025pathrag} ($\vs, \gs$): refine queries using keywords generated by gpt-4o-mini, prior to executing $\vs$ and $\gs$; and (4) \textbf{HippoRAG 2} ($\vs, \gs$): a KG-based method that we adapted to perform retrieval on corpus graph topology.
Finally, our \textbf{GraFine} ($\vs, \vgs, \gmodel$): utilizes frozen \texttt{[CLS]} features extracted from the reranker (pre-MLP) to initialize $\mathbf{H}$, leverages the last two layers of its classification head as $MLP(\cdot)$, and sets $b_\text{max}\!=\!100, BATCH\!=\!10, \alpha\!=\!0.2$, and $\beta\!=\!1$.
 
\subsubsection{RAG Baselines}
We generate responses by feeding the nodes retrieved by each retriever in the retrieval experiment into OpenAI's GPT-5-nano model. 
Consequently, the RAG experiment includes a total of 10 methods, nine baselines and GraFine.

\subsection{Datasets}
\label{subsec:datasets}

\subsubsection{The ACL-OCL dataset}
\label{subsec: acl-ocl}
ACL OCL~\cite{rohatgi-etal-2023-acl} is a text corpus derived from ACL Anthology, comprising approximately 80k academic papers with references and full texts. 
To evaluate baseline models on reference networks, we transform this corpus into the \textit{ACL-OCL dataset}, which will be publicly released, specifically designed to assess both retrieval and generation performance. 
Unlike existing datasets, ACL-OCL emphasizes scenarios where retrieving the correct answer requires understanding the semantic and structural context of intermediate nodes.
We construct the dataset through two stages: \textit{reference network construction} and \textit{synthetic query generation}.

\vspace{1mm}
\noindent \textit{Reference Network Construction.}
Each paper is divided into chunks of 4,096 characters, which constitute the node set $\mathcal{N}$. 
To construct the edge set $\mathcal{E}$ while mitigating spurious edges caused by paper-level metadata, we employ a reference detection model that identifies explicit citations within each chunk $n$.
For each detected citation, we create an edge $(n, n_i)$ linking the chunk to the cited paper’s corresponding node(s).
This model is implemented using GPT-5-nano via in-context learning. 
Detailed statistics are in Table~\ref{tab:dataset_stats}.

\vspace{1mm}
\noindent \textit{Synthetic Query Generation.} 
We generate synthetic query-gold node pairs by selecting connected node pairs and prompting an LLM to formulate questions requiring information from both.
This design intentionally targets the evaluation of retrieval-time refinement, as answering these queries necessitates interpreting node contents to bridge the semantic gap between the query and the target answer.
In total, we generated 753 queries, as illustrated in Figure~\ref{fig:acl-ocl-example}.

\begin{figure}[ht]
\small
\begin{tcolorbox}[
    colback=gray!5!white, 
    colframe=gray!75!black, 
    arc=1mm, 
    boxrule=0.5pt, 
    left=2mm, right=2mm, top=1mm, bottom=1mm 
]
    \textbf{\normalsize Query}  \vspace{1mm} \hrule \vspace{2mm}
    
    What visualization approach and export formats does the web-based annotation tool mentioned in \textbf{LIDA} use to render complex, overlapping text annotations and produce figures for publications?
    \vspace{2mm}
    
    \textbf{\normalsize Related Nodes (Oracle Nodes)}  
    \vspace{1mm} \hrule 
    \vspace{1mm}
    
    \textit{\textbf{LIDA node chunk \#2:}} \\
    \textcolor{red!80!black}{\textbf{BRAT (Stenetorp et al., 2012)}} and Doccano 3 are web-based annotation tools [\dots]. \textbf{LIDA} aims to fill these gaps by providing [\dots] 
    
    \vspace{2mm}
    \textit{\textbf{BRAT node chunk \#1:}} \\
    BRAT is based on our previously released opensource STAV text annotation visualiser[\dots] Both tools share a \textcolor{navyblue}{\textbf{vector graphics-based visualisation}} component [\dots] \textcolor{navyblue}{\textbf{BRAT integrates PDF and EPS image format}} export [\dots] 
    
\end{tcolorbox}
\vspace{-0.5cm}
\Description{Example of synthetic query generation. Red indicates textual reference to the BRAT node, while blue indicates the answer.}
\caption{Example of synthetic query generation. Red indicates textual reference to the BRAT node, while blue indicates the answer.}
\vspace{-0.1cm}
\label{fig:acl-ocl-example}
\end{figure}

\subsubsection{Datasets for Experiments}
To evaluate retrieval performance, we use a total of five datasets spanning two types of corpus graphs: ACL-OCL and LACD~\cite{an-etal-2025-grex} for reference networks, and BSARD-G, SciFact-G, and NFcorpus-G for text-rich knowledge graphs. 
As there are currently no established IR benchmarks specifically designed for text-rich knowledge graphs, we adapt three widely used IR benchmarks—BSARD \cite{bsard}, SciFact \cite{scifact-dataset}, and NFcorpus \cite{nfcorpus}—into graph-based formats following the graph construction procedure used in LightRAG. 
For ground-truth relevance, we define all nodes constructed from the original gold documents as oracle nodes.

For the RAG experiment, we use ACL-OCL for reference networks and two datasets from \textit{UltraDomain}~\cite{10.1145/3696410.3714805} for text-rich knowledge graphs. 
We excluded LACD, SciFact and NFCorpus from the RAG evaluation, as they are not formatted as QA datasets. 
Detailed statistics for all datasets are summarized in Table~\ref{tab:dataset_stats}.
\begin{table}[ht]
\centering
\vspace{1mm}
\caption{Dataset statistics for retrieval and RAG experiments. UD: UltraDomain datasets; Gray shading: \textit{reference networks}.}
\vspace{-0.35cm}
\label{tab:dataset_stats}
\renewcommand{\arraystretch}{0.8}
\resizebox{\columnwidth}{!}
{
\begin{tabular}{lccccc}
\toprule
\textbf{Datasets} & \textbf{Purpose} & \textbf{Domain} & $|\mathcal{N}|$ & $|\mathcal{E}|$ & $|\mathcal{Q}|$ \\
\midrule
\rowcolor{gray!30}
ACL-OCL     & Ret. \& Gen.  & CS          & 402,742   & 5,840,449 & 753\\
\rowcolor{gray!30}
LACD        & Retrieval     & Legal         & 192,974   & 339,666   & 89 \\
BSARD-G     & Ret. \& Gen.  & Legal         & 56,728    & 92,672    & 222 \\
SciFact-G   & Retrieval  & Science     & 36,438    & 43,557    & 1,109 \\
NFcorpus-G  & Retrieval     & Medical     & 23,468    & 27,805    & 3,237 \\
UD-agriculture & Generation & Agriculture & 46,561    & 82,088    & 100 \\
UD-mix      & Generation    & Mix         & 11,812    & 10,384    & 130 \\
\bottomrule
\end{tabular}
}
\vspace{-0.4cm}
\end{table}

\subsection{Evaluation Metrics}
\subsubsection{Conventional Metrics}
\label{subsubsec:metrics}

To evaluate retrieval performance, we use Capped Recall score ($R@k$) and Normalized Discounted Cumulative Gain (nDCG), both evaluated at \textbf{top-10}.
Capped Recall~\cite{thakur2021beir} normalizes the maximum achievable recall to 1 when selecting the top-$k$ nodes. 
For generation evaluation, we adopt a pairwise LLM-as-a-Judge approach, following the evaluation protocol and prompts used in LightRAG~\cite{guo2025lightrag}, where a generative LLM serves as the evaluator.

\subsubsection{Topological Recall}
\label{subsubsec:tr}

To quantify the effectiveness of \textit{retrieval-time refinement}, we introduce \textbf{Topological Recall (TR)}, a new metric defined over the range of $[0,1]$.
Unlike conventional Recall, TR captures the topological proximity of oracle nodes to the retrieved set $\mathcal{N}_\text{ret}$ by modeling shortest-path \textit{uncertainty}.

\begin{definition}[Topological Recall]
    For given $\mathcal{E}, \mathcal{N}_\text{ret}, \mathcal{N}_\text{oracle}$:
\[
TR(\mathcal{E}, \mathcal{N}_\text{ret}, \mathcal{N}_\text{oracle})=\mathrm{avg}_{n_i\in \mathcal{N}_\text{oracle}}(\frac{1}{1+u(\mathcal{E}, \mathcal{N}_\text{ret}, n_i)})
\]
where the uncertainty function $u$ is defined as the accumulated log-degree along the shortest path:
\[
u(\mathcal{E}, \mathcal{N}_\text{ret}, n_i) = \min_{n_j \in \mathcal{N}_\text{ret}} \sum_{n_k \in \mathrm{SP}(\mathcal{E}, n_j, n_i), n_k \ne n_i} \ln(1 + \mathrm{deg}_{\mathcal{E}}(n_k))
\]
Here, $\mathrm{SP}(\mathcal{E}, n_j, n_i)$ denotes the set of nodes on the shortest path from the seed node $n_j$ to the oracle node $n_i$.
\end{definition}

The intuition is that TR extends conventional Recall to capture the \textit{refinement potential} of a retrieved set. An unexplored oracle node is considered highly accessible if it is structurally close to the intermediate $\mathcal{N}_\text{ret}$ via low-degree neighbors. Conversely, high-degree nodes increase branching uncertainty, making it more likely that iterative \textit{adding} will introduce many irrelevant candidates that make it harder for iterative \textit{pruning} step to maintain well-\textit{refined} retrieved set. Thus, by penalizing the log-degree, TR measures how well the current node set is positioned to uncover oracle nodes via refinement.

To formalize this relationship and provide a theoretical foundation for its application in future research, we present the following decomposition as a corollary along with its proof.

\begin{corollary}[Decomposition]
    For given $\mathcal{E}, \mathcal{N}_\text{ret}, \mathcal{N}_\text{oracle}$, TR and Recall for $\mathcal{N}_\text{ret}$ have the following relationship:
    \[
TR=Recall+\frac{|\mathcal{N}_\text{oracle}\setminus \mathcal{N}_\text{ret}|}{|\mathcal{N}_\text{oracle}|}\cdot TR(\mathcal{E}, \mathcal{N}_\text{ret}, \mathcal{N}_\text{oracle}\setminus \mathcal{N}_\text{ret})
    \]
\end{corollary}
\begin{proof}
    Let $\mathcal{N}_\text{found} = \mathcal{N}_\text{oracle} \cap \mathcal{N}_\text{ret}$ and $\mathcal{N}_\text{miss} = \mathcal{N}_\text{oracle} \setminus \mathcal{N}_\text{ret}$.
    Note that for any $n_i \in \mathcal{N}_\text{found}$, the uncertainty $u(\mathcal{E}, \mathcal{N}_\text{ret}, n_i) = 0$.
    By decomposing the summation in the definition of TR:
    \begin{align*}
    TR &= \frac{1}{|\mathcal{N}_\text{oracle}|} \left( \sum_{n_i \in \mathcal{N}_\text{found}} 1 + \sum_{n_i \in \mathcal{N}_\text{miss}} \frac{1}{1+u(\mathcal{E}, \mathcal{N}_\text{ret}, n_i)} \right) \\
    &= \underbrace{\frac{|\mathcal{N}_\text{found}|}{|\mathcal{N}_\text{oracle}|}}_{\text{Recall}} + \frac{|\mathcal{N}_\text{miss}|}{|\mathcal{N}_\text{oracle}|} \underbrace{\left( \frac{1}{|\mathcal{N}_\text{miss}|} \sum_{n_i \in \mathcal{N}_\text{miss}} \frac{1}{1+u(\mathcal{E}, \mathcal{N}_\text{ret}, n_i)} \right)}_{TR(\mathcal{E}, \mathcal{N}_\text{ret}, \mathcal{N}_\text{miss})} \qedhere
    \end{align*}
\end{proof}

Hereafter, we refer to $\frac{|\mathcal{N}_\text{oracle}\setminus \mathcal{N}_\text{ret}|}{|\mathcal{N}_\text{oracle}|}\cdot TR(\mathcal{E}, \mathcal{N}_\text{ret}, \mathcal{N}_\text{oracle}\setminus \mathcal{N}_\text{ret})$ as \textit{MissTR}, meaning the partial credit for missing oracle nodes.
In Section~\ref{subsec:mechanism-analysis}, we analyze GraFine's capability for retrieval-time refinement using TR and MissTR.

%% file: sec5_result-and-analysis.tex
\section{Experimental Results}
\label{sec:result-and-analysis}

\subsection{Retrieval and Generation Performance}

\subsubsection{Retrieval Experiments}
Table \ref{tab:retrieval} presents the performance of GraFine and nine baselines for the retrieval pipeline across five graph retrieval datasets. 
Overall, our method demonstrates robust and consistent performance improvements across all evaluated datasets. 
\textbf{GraFine outperforms the most competitive baseline by 9.9\% in R@10 and 9.1\% in nDCG@10} in terms of overall average performance.

\begin{table*}[t]
\centering
\renewcommand{\arraystretch}{0.75}
\caption{Retrieval results on five corpus graph datasets. All metrics are reported in percentage (\%). The best results are highlighted in \textbf{bold}, and the second-best results are \underline{underlined}. \textit{Out-of-time} means that the method takes more than one hour per query.}
\vspace{-0.3cm}
\label{tab:retrieval}
\resizebox{\textwidth}{!}
{
\begin{tabular}{lrrrrrrrrrrrr}
\toprule
\multirow{4}{*}{\textbf{Methods}} & \multicolumn{4}{c}{\textbf{Reference Networks}} & \multicolumn{6}{c}{\textbf{Text-rich Knowledge Graphs}} & \multicolumn{2}{c}{\textbf{Average}} \\
\cmidrule(lr){2-5} \cmidrule(lr){6-11} \cmidrule(lr){12-13}
& \multicolumn{2}{c}{\textbf{ACL-OCL}} & \multicolumn{2}{c}{\textbf{LACD}} & \multicolumn{2}{c}{\textbf{BSARD-G}} & \multicolumn{2}{c}{\textbf{SciFact-G}} & \multicolumn{2}{c}{\textbf{NFCorpus-G}} & \multicolumn{2}{c}{\textbf{Overall}} \\
\cmidrule(lr){2-3} \cmidrule(lr){4-5} \cmidrule(lr){6-7} \cmidrule(lr){8-9} \cmidrule(lr){10-11} \cmidrule(lr){12-13}
& R@10 & nDCG@10 & R@10 & nDCG@10 & R@10 & nDCG@10 & R@10 & nDCG@10 & R@10 & nDCG@10 & R@10 & nDCG@10 \\
\midrule
\multicolumn{13}{l}{\textbf{Document Retrieval}} \\
Vector Search & 26.1 & 19.6 & 38.1 & 26.4 & 8.3 & 9.5 & 27.0 & 32.4 & 32.9 & 35.5 & 26.5 & 24.7 \\
SPLADE~\cite{formal2021splade} & \underline{39.2} & \underline{33.2} & 0.0 & 0.0 & 5.3 & 5.8 & 27.2 & 32.5 & 33.8 & 36.1 & 21.1 & 21.5 \\
Contriever~\cite{izacard2022unsupervised} & 20.4 & 16.0 & 1.1 & 0.9 & 0.4 & 0.3 & 27.7 & 32.5 & 35.2 & 37.7 & 17.0 & 17.5 \\
HyDE~\cite{gao-etal-2023-precise} & 22.7 & 17.1 & 39.2 & 26.3 & 9.9 & 11.5 & 29.8 & 35.7 & \underline{37.0} & \textbf{40.0} & 27.7 & 26.1 \\
Re2 & 29.6 & 24.3 & 47.9 & 33.1 & 10.8 & 12.2 & 29.4 & 34.7 & 34.6 & 37.1 & 30.5 & 28.3 \\
\midrule
\multicolumn{13}{l}{\textbf{Graph Retrieval}} \\
LightRAG~\cite{guo2025lightrag} & 24.2 & 15.7 & 19.3 & 12.0 & 11.0 & 10.2 & 31.4 & 32.0 & 34.3 & 35.4 & 24.0 & 21.1 \\
PathRAG~\cite{chen2025pathrag} & \multicolumn{2}{c}{\textit{Out-of-time}} & 0.0 & 0.0 & 11.3 & 11.7 & 14.4 & 14.8 & 31.6 & 33.4 & 14.3 & 15.0 \\
HippoRAG 2~\cite{gutierrez2025from} & 28.8 & 21.6 & 38.2 & 26.8 & 8.5 & 9.8 & 27.0 & 32.4 & 32.9 & 35.5 & 27.1 & 25.2 \\
GAR~\cite{macavaney2022adaptive} & 36.3 & 30.8 & \underline{48.6} & \underline{33.8} & \underline{12.8} & \underline{13.8} & \underline{32.5} & \underline{37.1} & 36.4 & 38.6 & \underline{33.3} & \underline{30.8} \\
\textbf{GraFine (Ours)} & \textbf{46.3} & \textbf{40.2} & \textbf{50.3} & \textbf{35.0} & \textbf{13.7} & \textbf{13.9} & \textbf{35.1} & \textbf{39.4} & \textbf{37.6} & \underline{39.3} & \textbf{36.6} & \textbf{33.6} \\
\bottomrule
\end{tabular}
}
\vspace{-0.3cm}
\end{table*}

\begin{table*}[htbp]
\centering
\caption{Overall Win Rates (\%) of Baselines v.s. GraFine across Four Datasets and Average.}
\label{tab:overall_win_rates_updated}
\vspace{-0.3cm}
\renewcommand{\arraystretch}{0.75}
\resizebox{\textwidth}{!}
{
\begin{tabular}{lcccccccccc}
\toprule
\multirow{2}{*}{\textbf{Baselines}} & \multicolumn{2}{c}{\textbf{ACL-OCL}} & \multicolumn{2}{c}{\textbf{BSARD-G}} & \multicolumn{2}{c}{\textbf{UltraDomain-agriculture}} & \multicolumn{2}{c}{\textbf{UltraDomain-mix}} & \multicolumn{2}{c}{\textbf{Average}} \\
\cmidrule(lr){2-3} \cmidrule(lr){4-5} \cmidrule(lr){6-7} \cmidrule(lr){8-9} \cmidrule(lr){10-11}
& Baseline & \textbf{GraFine} & Baseline & \textbf{GraFine} & Baseline & \textbf{GraFine} & Baseline & \textbf{GraFine} & Baseline & \textbf{GraFine} \\
\midrule
Vector Search & 45.6 & \textbf{53.3} & 42.3 & \textbf{57.2} & 43.0 & \textbf{55.0} & 40.8 & \textbf{57.7} & 42.9 & \textbf{55.8} \\
SPLADE & 48.2 & \textbf{51.7} & 27.0 & \textbf{72.5} & 43.0 & \textbf{57.0} & 45.4 & \textbf{54.6} & 40.9 & \textbf{59.0} \\
Contriever & 40.8 & \textbf{58.2} & 14.0 & \textbf{85.6} & 42.0 & \textbf{56.0} & 46.2 & \textbf{53.1} & 35.8 & \textbf{63.2} \\
HyDE & 44.1 & \textbf{55.4} & 48.6 & \textbf{51.4} & 44.0 & \textbf{56.0} & 41.5 & \textbf{58.5} & 44.5 & \textbf{55.3} \\
Re2 & 43.7 & \textbf{55.1} & 41.9 & \textbf{57.7} & 3.0  & \textbf{95.0} & 49.2 & \textbf{50.8} & 34.5 & \textbf{64.6} \\
\midrule
LightRAG & 42.5 & \textbf{57.1} & 45.9 & \textbf{53.6} & 39.0 & \textbf{60.0} & 45.4 & \textbf{53.1} & 43.2 & \textbf{55.9} \\
PathRAG & \multicolumn{2}{c}{\textit{Out-of-time}} & 42.8 & \textbf{56.8} & 38.0 & \textbf{62.0} & 23.1 & \textbf{76.2} & 34.6 & \textbf{65.0} \\
HippoRAG 2 & 45.8 & \textbf{53.5} & 43.2 & \textbf{56.8} & 46.0 & \textbf{53.0} & 35.4 & \textbf{63.1} & 42.6 & \textbf{56.6} \\
GAR-RAG & \textbf{52.1} & 47.1 & 48.2 & \textbf{51.8} & 38.0 & \textbf{61.0} & 39.2 & \textbf{60.0} & 44.4 & \textbf{55.0} \\
\bottomrule
\end{tabular}%
}
\vspace{-0.3cm}
\end{table*}

Specifically, GraFine significantly outperforms Re2, the strongest document retrieval baseline, by an average of \textbf{20.0\% in R@10 and 18.7\% in nDCG@10}.
Our method also yields substantial gains over GAR, the most competitive graph-based baseline, particularly in reference networks. For instance, in the ACL-OCL dataset, GraFine outperforms GAR by a relative margin of \textbf{27.5\% in R@10 and 30.5\% in nDCG@10}. Notably, on this same dataset, PathRAG fails to complete the retrieval within a reasonable time limit due to the high complexity of its flow algorithm. 
These results collectively highlight the effectiveness of our approach in navigating complex graph structures.

\subsubsection{RAG Experiments}
\begin{figure}[htb!]
    \centering
    \includegraphics[width=0.85\linewidth]{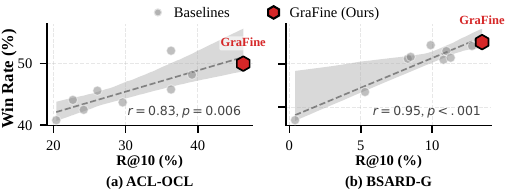}
    \vspace{-0.3cm}
    \Description{Correlation between R@10 and Win Rate. GraFine is the self-reference baseline (50\% win rate). Dashed lines and grey areas denote linear regression fits and 95\% CIs.}
    \caption{Correlation between R@10 and Win Rate. GraFine is the self-reference baseline (50\% win rate). Dashed lines and grey areas denote linear regression fits and 95\% CIs.}
    \label{fig:correlation}
    \vspace{-0.4cm}
\end{figure}

Table~\ref{tab:overall_win_rates_updated} presents overall win rates comparing baselines and our method over four datasets.
Here, GraFine demonstrates superior performance in the RAG setting, consistently achieving \textit{average win rates} \textbf{exceeding 55\% against all baselines.}
To investigate this improvement, we analyze the Pearson correlation between retrieval accuracy (R@10) and generation quality (win rate) on the ACL-OCL and BSARD-G datasets, as they are the only ones that support both retrieval and generation tasks.
As shown in Figure~\ref{fig:correlation}, we observe a significantly strong positive correlation with $p<0.05$.
It suggests that the retrieval enhancements from our method translate into more effective generation, indicating that employing GraFine as a retriever strengthens the overall Graph RAG capability.

\subsection{Efficiency and Latency Analysis}
\subsubsection{Query Processing Time Analysis}
\label{subsubsec:query-processing-time}
\begin{figure}[htb]
    \centering
    \includegraphics[width=0.85\linewidth]{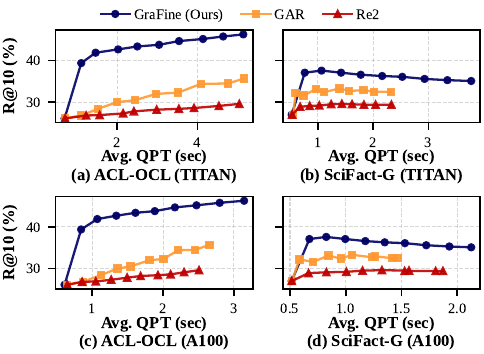}
    \vspace{-0.3cm}
    \Description{Scatter plots illustrating the trade-off between Average QPT and R@10 on (a,c) ACL-OCL, and (b,d) SciFact-G.}
    \caption{Scatter plots illustrating the trade-off between Average QPT and R@10 on (a,c) ACL-OCL, and (b,d) SciFact-G.}

    \label{fig:query_time}
\end{figure}

To demonstrate time-efficiency, we compare the \textit{Query Processing Time (QPT)} and R@10 of GraFine against baselines Re2 and GAR across TITAN and A100 GPUs. 
Figure~\ref{fig:query_time} illustrates the QPT and R@10 trade-off on ACL-OCL and SciFact-G. 
Each data point represents a different number of $\model$ operators ($b_{\max}$), from 10 to 100. 
As shown by the curves, GraFine achieves a \textbf{Pareto improvement}, consistently delivering higher R@10 without compromising efficiency on both datasets.

These results are also aligned with our discussion in Section~\ref{subsec:complexity-analysis}. 
Specifically, under the same budget \(b_{\max}=100\), the runtime of GraFine on ACL-OCL is comparable to that of GAR and Re2. 
Given that ACL-OCL contains 11$\times$ more nodes, 134$\times$ more edges, and 12$\times$ more edges per node than SciFact-G, this result suggests that the overall graph size is not the primary bottleneck of our method.
Moreover, on both ACL-OCL and SciFact-G, the latency of GraFine increases linearly as the budget grows. 
This observation supports our complexity analysis in Section~\ref{subsec:complexity-analysis}, indicating that the additional linear algebra computations are not the dominant bottleneck. 

\subsubsection{GraFine versus Conventional Interleaving Retrieval}
To demonstrate GraFine's efficiency over computationally intensive interleaving retrieval, we compare it against IRCoT + Vector Search~\cite{trivedi2023interleaving} and ToG-2~\cite{ma2025thinkongraph} on SciFact-G. 
To examine QPT fairly, we use a locally hosted Gemma 3 (12B) via Ollama in TITAN and A100 GPUs. 
IRCoT is configured with a 2-step process, retrieving 5 nodes per step.

\begin{table}[b!]
\centering
\renewcommand{\arraystretch}{0.8}
\caption{Time efficiency on SciFact-G: GraFine vs. LLM interleaving retrieval methods.}
\vspace{-0.3cm}
\label{tab:ircot}
\resizebox{\columnwidth}{!}{
\begin{tabular}{lccc}
\toprule
\textbf{Method} & \textbf{R@10} & \textbf{QPT (TITAN)} & \textbf{QPT (A100)} \\
\midrule
IRCoT + Vector Search & 31.4 & 6.54 sec & 5.03 sec \\
ToG-2 & 12.3 & 32.3 sec & 20.7 sec \\
\textbf{GraFine (Ours)} & \textbf{35.1} & \textbf{3.77 sec} & \textbf{2.12 sec}\\
\bottomrule
\end{tabular}
}
\end{table}

As shown in Table~\ref{tab:ircot}, IRCoT slightly improves over Vector Search in terms of accuracy, but incurs substantial latency due to iterative LLM inference. 
ToG-2 further amplifies this limitation, exhibiting significantly higher latency with degraded retrieval performance.
In contrast, GraFine effectively overcomes these bottlenecks. 
Compared to IRCoT, our method reduces \textbf{query processing time by 42--58$\mathbf{\%}$ while improving R@10 by 11.7\%}, and achieves both \textbf{higher accuracy and around 10$\times$ faster latency} than ToG-2, validating it as a time-efficient alternative to interleaving retrieval methods.





\subsection{Understanding GraFine}
\label{subsec:mechanism-analysis}

\subsubsection{Topological Recall (TR) Analysis}
In this section, we utilize the Topological Recall (TR) metric defined in Section~\ref{subsubsec:tr} to analyze how effectively GraFine refines the retrieved node set.
We validate whether TR effectively quantifies the topological proximity to oracle nodes and how our method exploits this proximity relative to baselines.

\begin{figure}[tb]
    \centering
    \includegraphics[width=0.85\linewidth]{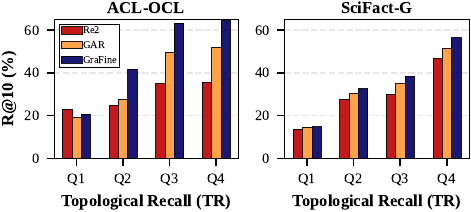}
    \vspace{-0.4cm}
    \Description{Impact of Topological Recall (TR) on Retrieval Performance (R@10).}
    \caption{Impact of Topological Recall (TR) on Retrieval Performance (R@10).}
    \vspace{-0.1cm}
    \label{fig:tr_recall_histogram}
\end{figure}

\vspace{1mm}
\noindent \textit{Impact of Topological Proximity.} We first investigate the relationship between TR and the standard Recall (R@10) to understand how topological proximity translates to retrieval performance. Figure \ref{fig:tr_recall_histogram} illustrates the distribution of R@10 across TR quartiles (Q1--Q4). As expected, all methods demonstrate improved Recall with increasing TR, confirming that higher topological proximity creates a favorable state for uncovering remaining oracle nodes through refinement.
Critically, graph traversal methods (GraFine and GAR) exhibit steeper performance gains in the Q3-Q4 intervals compared to the vector-only baseline, Re2.
This indicates that graph-based methods successfully exploit the graph structure to retrieve oracle nodes that remain unreachable through vector similarity alone.
In particular, GraFine outperforms Re2 by a larger margin than GAR in the structurally difficult Q2 interval of ACL-OCL, demonstrating its ability to effectively navigate towards oracle nodes even when the initial search starts in a poorly connected region of the graph.

\vspace{1mm}
\noindent \textit{Correlation Analysis.} To substantiate these observations, we examine which component of TR captures topological proximity to oracle nodes. Based on Corollary 1, we analyze the correlation between the two components–\textit{Recall} and \textit{MissTR}–against the \textit{marginal recall gain} ($\Delta R =R@100_{total} - R@10_{vs}$ where $R@100_{total}$ is the recall after complete retrieval and $R@10_{vs}$ is the recall obtained from initial \vs). Here, $\Delta R$ quantifies the retriever's success in uncovering undiscovered oracle nodes during graph retrieval.

\begin{table}[ht]
\centering
\caption{Correlation coefficient between two metrics and $\Delta R$.}
\vspace{-0.3cm}
\label{tab:gq-correlation}
\renewcommand{\arraystretch}{0.85} 
\resizebox{\columnwidth}{!}
{
\begin{tabularx}{\columnwidth}{l*{6}{>{\centering\arraybackslash}X}}
\toprule
\multirow{2}{*}{\textbf{Methods}} & \multicolumn{2}{c}{\textbf{ACL-OCL}} & \multicolumn{2}{c}{\textbf{BSARD-G}} & \multicolumn{2}{c}{\textbf{SciFact-G}} \\
\cmidrule(lr){2-3} \cmidrule(lr){4-5} \cmidrule(lr){6-7} 
 & Recall & MissTR & Recall & MissTR & Recall & MissTR \\
\midrule
GAR & 0.48 & \textbf{0.50} & 0.46 & \textbf{0.79} & 0.12 & \textbf{0.65} \\
GraFine & 0.65 & \textbf{0.66} & 0.32 & \textbf{0.57} & 0.10 & \textbf{0.55} \\
\bottomrule
\end{tabularx}
}
\end{table}

As shown in Table \ref{tab:gq-correlation}, MissTR consistently exhibits a stronger correlation with $\Delta R$ than standard Recall across all datasets and methods. While Recall reflects the success of the initial $\vs$ retrieval, it shows weak correlation with future discoveries. 
In contrast, the strong correlation in MissTR suggests that this component effectively captures the \textit{refinement potential} of the current node set $\mathcal{N}_\text{sel}$. These findings support our hypothesis that performance gains in graph-based methods are driven by their ability to navigate toward undiscovered oracle nodes through the graph topology.

\begin{figure}[htb]
    \centering
    \includegraphics[width=0.85\linewidth]{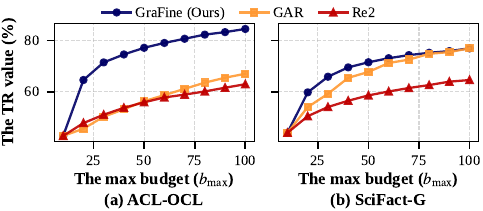}
    \vspace{-0.4cm}
    \Description{Evolution of Topological Recall (TR) as a function of retrieval budget ($b_{max}$)}
    \caption{Evolution of Topological Recall (TR) as a function of retrieval budget ($b_{max}$)}
    \vspace{-0.1cm}
    \label{fig:gq_performance}
\end{figure}

\noindent \textit{Evolution of TR in retrieval.} Figure \ref{fig:gq_performance} illustrates the evolution of TR as the retrieval budget $b_{max}$ increases. 
While all methods exhibit an upward trend, the rise in Re2 is largely a trivial consequence of the increasing Recall term within the decomposed TR equation (Corollary 1).
More critically, we distinguish the trajectories of the graph traversal methods. Unlike the gradual ascent of GAR,
GraFine demonstrates a rapid initial surge 
in TR. This sharp trajectory validates the efficacy of our proposed operators:
SPX actively \textit{steers} node selection towards oracle-rich neighborhoods in the early retrieval stages, while GSR effectively prioritizes candidates by interpreting their topological context. 
Consequently, these results provide strong empirical evidence that GraFine's mechanisms support faster retrieval-time refinement toward high topological proximity than competing approaches.

\begin{figure}[tb!]
    \centering
    \includegraphics[width=0.85\linewidth]{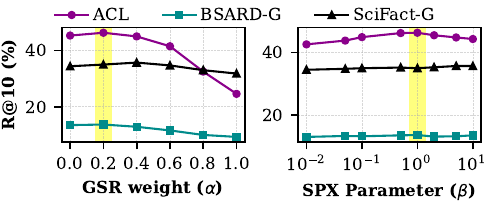}
    \vspace{-0.4cm}
    \Description{
    R@10 sensitivity to GSR weight (left) and SPX parameter (right). Yellow bands mark the default values.
    }
    \caption{
    R@10 sensitivity to GSR weight (left) and SPX parameter (right). Yellow bands mark the default values.
    }
    \vspace{-0.1cm}
    \label{fig:parameter-sensitivity}
\end{figure}

\subsubsection{Sensitivity Analysis} 
\label{subsubsec:hyperparam-sensitivity}
Figure~\ref{fig:parameter-sensitivity} illustrates the sensitivity of GraFine's R@10 performance to variations in hyperparameters $\alpha$ and $\beta$ across three datasets. 
The results show consistently high performance
across all datasets at our chosen settings of $\alpha=0.2$ and $\beta=1$, justifying our parameter selection.
Conversely, we observe a performance degradation when $\alpha$ approaches 0 (i.e., relying solely on $\model$ rather than $\gmodel$) or when $\beta$ approaches extreme values of 0 or $\infty$ (i.e., using only $\vs$ or $\gs$ instead of $\vgs$). 
These findings demonstrate the contribution of both GSR and SPX to the overall effectiveness of our proposed method.

%% file: sec6_related-works.tex
\section{Related Work}
\label{sec:related-works}

Recently proposed retrieval--LLM interleaving methods for retrieval-time refinement~\cite{sun2024thinkongraph,ma2025thinkongraph, trivedi2023interleaving, mavromatis2024gnn, lee-etal-2024-planrag}, while effective for problems that go beyond single-step retrieval, rely on frequent LLM invocations, which incur substantial computational overhead and latency, making them impractical for Graph RAG on corpus graphs.
Other recent graph retrieval methods aim to perform effective retrieval-time refinement by combining two or more operators from $\{\vs, \gs, \model\}$~\cite{guo2025lightrag, chen2025pathrag, jimenez2024hipporag, gutierrez2025from}.
However, these methods are fundamentally based on a sequential composition of operators and therefore inherit the limitations of those operators---the topology blindness of $\model$ and the semantic blindness of $\gs$.
Moreover, we observe that existing \textit{fusion} approaches, such as G-Retriever~\cite{he2024g}, can also be formally expressed as compositions of our proposed operators $\{\vs, \gs, \model\}$. Specifically, Examples~\ref{eg:ppr} and~\ref{eg:g-retriever} demonstrate how the PPR algorithm in HippoRAG 2~\cite{gutierrez2025from} and G-Retriever~\cite{he2024g} can be represented within our taxonomy, respectively.

%% file: sec7_conclusion.tex
\section{Conclusion}
\label{sec:conclusion}

In this paper, we presented \textbf{GraFine}, a novel graph retrieval method designed to enable time-efficient retrieval-time refinement for Graph RAG on corpus graphs.
Specifically, we identify two blind points of existing retrieval operations and overcome them by interleaving two novel fusion operators: \textbf{the Graph Smoothing Reranker (GSR)} for $\gmodel$ and \textbf{Semantic Proximity eXpansion (SPX)} for $\vgs$.
Extensive experiments across five corpus graph datasets demonstrate that GraFine outperforms state-of-the-art baselines by an average of \textbf{+9.9\% in R@10 and +9.1\% in nDCG@10.} Furthermore, compared to conventional LLM interleaving methods, our approach achieves a significant Pareto improvement, \textbf{reducing query processing time by 42--58\% while simultaneously improving R@10 by 11.7\%.}

%% file: sec8_genAI.tex
\section*{GenAI Usage Disclosure}
Generative AI were used in a limited manner during the preparation of this work, including linguistic assistance during drafting, code debugging assistance during implementation, certain stages of ACL-OCL dataset construction (Section~\ref{subsec: acl-ocl}), and the LLM-as-a-Judge evaluation (Section~\ref{subsubsec:metrics}). 
Except for the linguistic assistance and code debugging, all GenAI-assisted components are described in detail in the paper. 
No generative AI tools were used in any other part of the research or writing process.